\documentclass[showpacs,preprint,amssymb]{revtex4}
\usepackage[dvips]{epsfig}
\usepackage{float}
\usepackage{graphicx}
\usepackage{dcolumn}
\usepackage{bm}
\def\be{\begin{equation}} 
\def\ee{\end{equation}}
\def\bea{\begin{eqnarray}} 
\def\eea{\end{eqnarray}}
\def\line{\hbox to \hsize}    
\def\frac #1#2{{#1\over #2}}

\def \ket #1{{\vert #1\rangle}}
\def \bra #1{{\langle #1\vert}}

\def\1{\mbox{\bf 1}}
\newcommand{\comment}[1]{}

\begin{document}

\title{On the $Z_2$  classification of  Quantum Spin Hall Models}
\author{Rahul Roy}
\affiliation{University of Illinois, Department of Physics\\ 1110 W. Green St.\\
Urbana, IL 61801 USA\\E-mail: rahulroy@uiuc.edu}

\begin{abstract}
 We propose an alternative formulation of the $Z_2$ topological index for quantum spin Hall systems and band insulators when time reversal invariance is not broken. The index is expressed in terms of the Chern numbers of the bands of the model, and a connection with the number of pairs of robust edge states is thus established. The alternative index is  easy to compute in most cases of interest. We also discuss connections with the recently proposed spin Chern number for quantum spin Hall models. 
 \end{abstract}

\maketitle
     Two dimensional models with time reversal symmetry where momentum space topology leads to quantized spin Hall responses to external electric fields  have recently been proposed in the presence of Landau levels arising due to  a strain gradient \cite{BernevigandZhang} and in graphene and other systems \cite{Kaneetal1,Qietal1,RRoy}. Kane and Mele \cite{Kaneetal2} have recently introduced a $Z_2 $ classification for these quantum spin Hall models motivated by the  observation that two time reversed pairs of edge states could scatter with each other and localize but the same cannot be said of a single pair of edge states. They point out that results from twisted real K theory \cite{Atiyah1} dictate the presence of a $Z_2 $ classification of spin Hall ground states and  propose an index, $I$ which counts the number of points in the Brillouin zone that belong to the ``odd subspace" modulo two.
      
     Though the formulation of the index is for a clean non-interacting system, the topological invariance suggests robustness of the edge modes of the system against small disorder and weak interactions, a fact that seems to be borne out by simulations \cite{Shengetal}, though the gapless edge states are not stable to multiparticle scattering processes and strong interacions \cite{XuandMoore, Wuetal}. 
     Kane and Mele's index $I$ can be evaluated by counting the number of complex zeros of the Pfaffian function $P(k)$ given by
     \[
     P(k) = Pf[\bra{u_i ({\bf k})}\Theta\ket{u_j ({\bf k})}] 
     \]
     where $ \Theta$ is the time reversal operator and $u_i, u_j $ are the occupied Bloch wavefunctions in momentum space. 
     The index $I$ is found by evaluating the winding of the phase of $P(k)$ around a loop enclosing half the Brillouin zone such that only one of ${\bf k}$ and ${\bf -k}$ are enclosed for each ${\bf k}$, 
     \[
      I = {1 \over 2\pi i}\oint_C d{\bf k}. \nabla_{\bf k}\log(P(k))
     \]
     
   Here we propose an alternate $Z_2 $ index for time reversal invariant systems where the square of the time reversal operator is -1, which in the absence of disorder and interactions is related to the TKNN\cite{TKNN} numbers of the bands in the system. We are thus able to make a connection with the number of edge modes and gapless excitations in the system. 
 
      The Hall conductance of certain tightbinding lattice model is given by the sum of the Chern numbers of the filled bands in momentum space. If the bands touch, the bands may exchange Chern numbers but the total Chern number is conserved \cite{Avron}. The Hall conductance is thus quantized as long as there is a gap in the excitation spectrum. This argument can be extended to non interacting systems by expressing the Hall conductance as a Berry phase in the space of the parameters which can be used to specify  generalized periodic boundary conditions \cite{Niuetal}.   
   
   In systems with time reversal invariance, the bands come in time reversed pairs with opposite Chern numbers and the net Chern number is zero. In such systems the Hall conductance is zero, but a spin Hall response is possible. A number of recent lattice models have been proposed where a spin quantized response is obtained \cite{Kaneetal1,Qietal1,RRoy}. It turns out that these time reversed pairs of bands can exchange Chern numbers. However, there is a $Z_2 $ index which remains invariant under band mixing as long as the gap between the ground states and the excited states does not close. This is the main result of this paper. 
   
    The $Z_2 $ classification can be analysed in the general case when the ground state has $2p$ filled bands (see Appendix). For simplicity, we shall restrict ourselves to the case when $p=1$.
    A generic four band, time reversal invariant, tight binding Hamiltonian can be written in momentum space in the form \cite{Kaneetal2}: 
  \begin{eqnarray}
H= \sum_{a=1}^{5} d_{a}\Gamma_a + \sum_{a<b=1}^{5}d_{ab}\Gamma_{ij} 
\end{eqnarray}
     where the $\Gamma_a$'s are the matrices defined in \cite{Kaneetal2} and the $d_a$'s and $d_{ab}$'s are symmetric and antisymmetric functions in momentum space as required by time reversal invariance.  
     The ground state then consists of two bands which map onto each other under time reversal.   The topology of the Brillouin zone for a two dimensional periodic lattice is that of a two dimensional torus. The bands form a rank two vector bundle on the torus. Time reversal symmetry leads to an involution on the torus, and on the fibre bundle, the bundle is thus ``twisted" and can be classified using real K theory. They point out that results from twisted real K theory \cite{Atiyah1} dictate the presence of a $Z_2 $ classification of spin Hall ground states and  propose an index, $I$ which counts the number of points in the Brillouin zone that belong to the ``odd subspace" modulo two.
  
\begin{figure} 
  \includegraphics[scale=0.50]{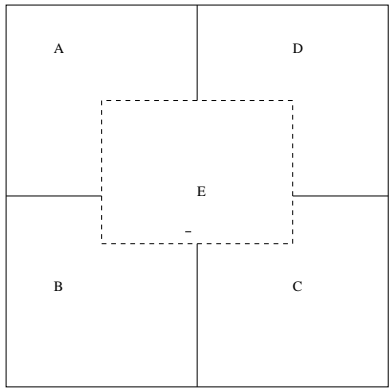}
  \caption{\label{fig1}The two dimensional torus divided into five regions, A, B,C,D and E. The dashed curve marks the path along which we study the transition matrices }
  \end{figure}

   Consider a two dimensional torus which we divide into five non overlapping regions A,B,C,D and E as shown in Fig. \ref{fig1}. We take A', B', C', D' and E' as regions slightly larger than these regions and study the transition matrices in the overlaps of these regions. 
  
  We are only interested in the topological distinctions between ground states that affect the spin Hall response and are connected to the number of edge modes in the system. Thus we are interested in properties associated with the curvature of the Berry phase. 
   For our purposes, it is sufficient to study the set of vector bundles which are homotopic to the one whose transition matrices across the regions, A', B' etc. differs from the identity only along the dashed line marked in the figure. We define the line bundles $ (1,0)^T $ and $(0,1)^T$ on the two regions separated by the dashed curve, (namely the region E and everything outside ) such that the operation of time reversal on an arbitrary vector in these regions corresponds to the operator $i\sigma_2 K_0 \phi$ where $K_0$ is the complex conjugation operator and $\phi$ is the involution operator that takes vectors at ${\bf k}$ to vectors at ${\bf - k}$. We parametrize the dashed curve by the angle $\phi \in S^1$, (where $S^1$ is the unit circle), which varies from $0$ to $2\pi$. If we now write down the transition matrix $U(\phi) \in U(2)$ in the form $ e^{i\alpha(\phi)I}e^{i{\bf n(\phi).\sigma} \omega(\phi)} $, then time reversal invariance leads to the condition  that 
  \bea \label{tricondition}
  U(\phi+\pi) = e^{-i\alpha(\phi)I}e^{i{\bf n(\phi).\sigma} \omega(\phi)}
  \eea
  
  One can then show (see Appendix) that the set of maps  that satisfy the above condition can be continuously transformed (along with a suitable gauge transformation) either to the constant map $ U(\phi)= I$  or to the map $ U(\phi) = e^{-i{\pi\over 2}I}e^{i(\phi+{\pi\over 2})\sigma_3} $ (see. When the bands are distinct, the Chern numbers of the bands correspond to the winding numbers of the diagonal elements. These then fall into two categories : when the bands carry even Chern numbers, the ground state is topologically equivalent, ( i.e may be adiabatically continued) to the state with no edge modes and, when the bands carry odd Chern numbers, the ground state has atleast one pair of edge modes. In a recent paper \cite{Qietal2} the connection between the bulk topology and gapless excitations at the edges in various systems is discussed.
  
    A similar analysis can be carried out for the case when $p\neq 1$ (see Appendix)
  Thus we are led to define an alternative index, $ E= 0$ for the first case and $E=1$ for the second. 
   
    If we then define \[ C = \sum_{c_n>0 } c_n \] where n is the band index and $c_n $ is the corresponding Chern number and the sum is taken over the set of bands which have positive Chern numbers,  then
    \bea
     E = C \,\textrm{mod}\, 2
     \eea
    These results can also be understood from a band touching picture. Time reversal invariance dictates that when time reversed pairs of bands touch, they do so at an even number of points, and the Chern number exchange is always an even number. However, two sets of such bands, each with Chern numbers $\pm 1$, may touch with the Chern number of each band after the exchange being zero. 
    
    The extension of the TKNN\cite{TKNN} numbers to systems with interactions and disorder is usally done by expressing the Hall conductance of the bands in terms of the Chern numbers in the space of the parameters that determine generalized periodic boundary conditions. A spin Chern number has been recently proposed by extending this idea to spin quantum Hall systems\cite{Shengetal}. Our analysis suggests that while this analysis is useful for the case when the Hamiltonian of the system can be derived from a Hamiltonian which has  a pair of occupied bands whose Chern numbers are $\pm 1$ in the ground state, ( with the rest of the occupied bands having zero Chern number) it might fail when this is not the case. 

\begin{center}
APPENDIX
\end{center}

  Here we  elaborate on the homotopy arguments for the transition functions for both single and multiple pairs of bands.  \\
   It follows from Eq.[\ref{tricondition}] that if we write $U(\phi)$ as a function $ e^{i\theta(\phi)I}g(\phi)$ with  $ \theta(\phi) \in R , g(\phi) \in SU(2) $ being continuous functions, they satisfy  the condition
  $ \theta(\pi) = -\theta(\pi) + p \pi , g(\pi) = e^{-i p\pi I} g(0) $. If $p$ is even, $g(p) = g(0)$ and since 
  $ \pi_1 (SU(2))=0 $, it follows that the transition function in this case can be continuously deformed to the constant function, $ U(\phi) = e^{-ip\pi \over 2} g(0) $.     
     When $p$ is odd, then $ g(\pi ) = - g(0)$ and thus $ g(\pi) $ and $g(0)$ are always distinct elements for any value of $g(0)$. We can deform an arbitrary function $ g(\phi)$ which satisfies this condition to the particluar one, 
  $ e^{-ip\phi\over 2} e^{i(\phi+{\pi \over 2})\sigma_3 } $. The constants for both cases can be eliminated by means of a gauge transformation.  
   
   To make the connection with the Chern numbers of the bands, note that any transition function which corresponds to bands which have Chern numbers $ n, -n$ can be continuously deformed to the diagonal matrix with entries $ e^{i n\phi}, e^{-in\phi}$. It is readily verified that $p$ is equal to $\pm n$.
 
   If we have N pairs of bands, with the  basis in each region again chosen such that the time reversal operator is $ J K_0\phi $ where J is the matrix whose only nonzero elements are $ J(2i, 2i+1) = -J(2i+1,2i) = 1 $ and $K_0$ the complex conjugation operator as before, the corresponding condition for the continuous  functions $\theta(\phi), g(\phi) $ where the transition functions is  $U(\phi) = e^{i\theta(\phi)} g(\phi) $ is that 
   $ \theta (\pi) = -\theta + {p\pi \over N} , g(\pi) = e^{-i p\pi\over N} J g(0)J^{-1} $. In this case too there are two classes of transition functions corresponding to topologically equivalent band structures, depending on whether $p$ is odd or even. The even class consists of transition functions which can be continuously deformed to the constant function and the odd class to a a constant multiple of the matrix whose only non zero elements are $ U[1,1]\, (\phi) = e^{i\phi}, U[2,2]\,(\phi) = e^{-i(\phi+\pi)}, U[i,i] = 1, i \neq 1,2 $. In this case again, it is easy to verify that the two classes of vector bundles can be labeled by the $Z_2 $ index, E. 
 \\ 
 
   I would like to thank Prof. Sheldon Katz, Prof. Michael Stone and Prof. Shivaji Sondhi for useful and stimulating discussions and helpful suggestions and the University of Illinois Research Board for support. I would also like to thank Prof. Katz for explaining various aspects of Ref.\cite{Atiyah1} and twisted KR theory and Dr. S. K. Roy and Dr. Kumar Raman for valuable suggestions.
   
    I would also like to acknowledge Prof. Joel Moore and Prof. Leon Balents for their constructive reading of version 1 of this preprint \cite{version1} and their suggestions on including details of the topological arguments which now form the appendix. I would also like to draw the reader's attention to Ref.\cite{Moore2006top} which addresses the problem from a different point of view.

 \end{document}